\documentclass[showpacs,preprintnumbers,amsmath,amssymb]{revtex4}
\usepackage{graphicx,epsfig}

\def\Journal#1#2#3#4{{#1} {\bf #2}, #3 (#4)}

\def\PLB{{Phys. Lett.}  B}
\def\PLA{{Phys. Lett.}  A}
\def\PRL{\em Phys. Rev. Lett.}
\def\PRD{{Phys. Rev.} D}
\def\PR{{Phys. Rev.}}

\def\CQG{{Class. Quant. Grav.}}
\def\CMP{{Commun. Math. Phys.}}
\def\JMP{{J. Math. Phys.}}
\def\JPA{{J. Phys.}  A}
\def\IJMPA{{Int. J. Mod. Phys.} A}
\def\ibid{{\it ibid.}}

\newcommand{\be}{\begin{equation}}
\newcommand{\ee}{\end{equation}}
\newcommand{\bea}{\begin{eqnarray}}
\newcommand{\eea}{\end{eqnarray}}

\begin{document}

\title{Scalar hairy black holes and solitons in asymptotically flat spacetimes}

\date{\today}


\author{Ulises Nucamendi}
\email{ulises@fis.cinvestav.mx} \affiliation{Instituto de
F\'{\i}sica y Matem\'{a}ticas, Universidad Michoacana de San
Nicol\'{a}s de Hidalgo,\\
Edif. C-3, Ciudad Universitaria, Morelia, Michoac\'{a}n, C.P.
58040, M\'{e}xico}

\author{Marcelo Salgado}
\email{marcelo@nuclecu.unam.mx} \affiliation{Instituto de Ciencias
Nucleares, Universidad Nacional Aut\'onoma de M\'exico,\\
 A.P. 70-543, M\'exico 04510 D.F., M\'exico}

\begin{abstract}
A numerical analysis shows that a class of scalar-tensor 
theories of gravity with a scalar field minimally and nonminimally 
coupled to the curvature allows static and spherically symmetric 
black hole solutions with scalar-field 
hair in asymptotically flat spacetimes. In the limit when the 
horizon radius of the black hole tends to zero, regular scalar solitons 
are found. The asymptotically flat solutions are obtained provided 
that the scalar potential $V(\phi)$ of the theory 
is not positive semidefinite and such that its local minimum is also a zero of the 
potential, the scalar field settling asymptotically at that minimum. 
The configurations for the minimal coupling case, although unstable under spherically 
symmetric linear perturbations, are regular and thus can serve as 
 counterexamples to the no-scalar-hair conjecture. For the nonminimal coupling case, 
the stability will be analyzed in a forthcoming paper.
\end{abstract}

 \maketitle


It was more than thirty years ago when Ruffini and Wheeler \cite{ruffini}
proposed the so called no-hair conjecture for black holes. This conjecture
states that black holes are completely characterized by their mass, charge,
and angular momentum. In order to prove this conjecture, some people
established several no-hair theorems in theories which couple classical fields
to Einstein gravity, notably the black hole uniqueness theorems in
Einstein-Maxwell (EM) theory \cite{israel} which establishes that all the
solutions of black holes in EM theory are stationary and axially symmetric and 
contained within the Kerr-Newman family. Other theorems proved by
Chase \cite{chase}, Bekenstein \cite{bekenstein}, Hartle \cite{hartle} and
Teitelboim \cite{teitelboim}, showed that stationary black holes solutions are
hairless in a variety of theories coupling different classical fields to 
Einstein gravity. A key ingredient in the no-hair-theorem proofs relies 
on the assumption of asymptotic flatness (AF) and on the 
nature of the energy-momentum tensor. For instance, 
in recent years counterexamples to the no-hair conjecture
were found in several theories with non-Abelian gauge fields which include the
Einstein-Yang-Mills \cite{eym}, Einstein-Yang-Mills-Higgs (EYMH) 
\cite{eymh}, Einstein-Yang-Mills dilaton (EYMD) \cite{eymd}, 
the Einstein-Skyrme \cite{es} and 
the Einstein-non-Abelian-Procca \cite{eymh} theories. 
The existence of solitons and hairy black holes in these theories \cite{review}
is associated with the non-Abelian gauge fields present in them 
\cite{comment1}. 
In the case of matter composed by a single scalar field minimally 
coupled (MC) to gravity, Sudarsky \cite{sud1},
has proved a very simple no-hair theorem. Among other assumptions, 
the validity of that theorem relies on the fulfillment of the weak energy 
condition (WEC) which constrains the scalar potential to be positive 
semi-definite. Recently a more general no-hair theorem for black holes 
has been proved. This rules out a multicomponent scalar field coupled minimally to 
gravity satisfying the WEC, but its field
Lagrangian is not quadratic in the field derivatives 
\cite{bekenstein2}. In fact a central argument for probing all these 
theorems relies on the fulfillment of the WEC which is generally believed
to be satisfied for all physically reasonable classical matter
\cite{wald2}. 

In the case of nonminimally coupled (NMC) scalar fields, there are 
only a few no-hair theorems \cite{zannias,nomin}. The AF hairy black hole 
solution corresponding to a scalar field conformally coupled to gravity 
(the Bronnikov-Melnikov-Bocharova-Beckenstein solution \cite{BMBB}) 
which is often used as a counterexample for the failure of the 
no-scalar-hair conjecture, has been criticized as a non genuine black 
hole solution in that the Einstein field equations are not verified 
at the horizon \cite{sudzann}, and moreover, in that 
it is not a solution at all if one demands a bounded scalar field 
throughout the static region [12(a)]. 

Recently, in the context of asymptotically AdS spacetimes, scalar-hairy black 
holes (SHBH) were found for some scalar-field potentials $V(\phi)$ 
\cite{japs}. 
A further analysis along this line \cite{sudja}, suggested in addition that 
by adjusting the parameters of $V(\phi)$, the effective cosmological constant 
could cancel out so that asymptotically flat SHBH could exist.

Encouraged by the findings in the AdS context, we show in this letter 
numerical evidence of SHBH solutions in static, 
spherically symmetric and asymptotically flat 
spacetimes. The solutions are regular throughout the static region (from the 
Killing horizon to spatial infinity). In particular, for the MC case, 
the main assumption for the Sudarsky's no-hair theorem \cite{sud1} not to be 
valid here is that the class of scalar-field potentials $V(\phi)$ 
used to construct 
the solutions are not positive-semidefinite (the WEC is violated). 
Moreover, the class $V(\phi)$ is in fact  
such that $V(\phi_c)=0= \partial_\phi V|_{\phi_c}$ and 
$\partial^2_{\phi\phi} V|_{\phi_c} > 0$, that is, 
a root of $V(\phi)$ and a local minimum are located 
at the same place. This feature and the fact that 
the non-trivial scalar field settles asymptotically 
at the local minimum (and therefore also at 
a root) of $V(\phi)$, leads to configurations that truly represent 
asymptotically flat solutions. 
When taking the limit $r_h\rightarrow 0$ ($r_h$ being the horizon radius), 
the SHBH tend to regular scalar solitons (scalarons). 
Such results can extend to the NMC case as it turns numerically 
(the no-hair theorems for the NMC are avoided by considering the scalar 
potentials as mentioned above).

The SHBH solutions although genuine asymptotically flat solutions, 
and perhaps the first examples with a regular horizon and with an explicit
scalar field potential, constitute only a ``weak'' counterexample to the 
no-scalar-field-hair conjecture since they turn to be unstable with 
respect to radial perturbations.


{\it Einstein-scalar field equations and Lagrangian}.- We will
consider a model of a scalar field NMC to
gravity and with a potential. The simplest model of this kind is
obtained by considering the Lagrangian
\be
{\cal L} = \sqrt{-g} \left[ { 1\over 16\pi } R 
+ \xi\phi^2 R - {1\over 2}(\nabla \phi)^2 - V(\phi) \right]  
\,\,\,\,\,.
\label{lag} 
\end{equation}
By choosing the NMC constant $\xi=0$, the above theory corresponds to the minimal coupling case, 
and $\xi=-1/12$ corresponds to the conformally coupled scalar field 
(units where $G_0=c=1$ are employed). 
The gravitational field and scalar field equations 
following from the Lagrangian (\ref{lag})
can be written as
\be
G_{\mu\nu} = 8\pi T_{\mu\nu} \,\,\,,\,\,\,
\Box \phi + 2\xi R = \frac{\partial V(\phi)}{\partial\phi}\,\,\,\,,
\end{equation}
where 
\be
T_{\mu\nu} = G_{\rm eff} \left\{
(\nabla_\mu\phi)\nabla_\nu\phi
- g_{\mu\nu}\left[{1\over 2}  (\nabla \phi)^2 + V(\phi)\right]
+ 4\xi \left[\nabla_\mu(\phi\nabla_\nu\phi) - g_{\mu\nu}
\nabla_\lambda (\phi \nabla^\lambda\phi)\right]
\right\}\,\,\,,
\end{equation}
is an effective energy-momentum tensor which includes 
all the contributions of the scalar field, and $G_{\rm eff}$ 
is an effective gravitational ``constant'' which explicitly depends
on the scalar field: 
\be
G_{\rm eff} ={1\over (1 + 16\pi\xi \phi^2)}\ .
\label{geff}
\end{equation}

We will focus on a metric describing spherical and static
spacetimes: 
\be 
ds^2 = - N e^{2\delta} dt^2 + N^{-1} dr^2 + r^2
(d\theta^2 + \sin^2\theta d\varphi^2) \,\,\,\,\,, \label{RGMS} 
\end{equation}

where $N \equiv (1 - 2m(r)/r)$ and the functions $m$ and
$\delta$ depend only on the coordinate $r$. For the
scalar field we also assume $\phi=\phi(r)$. 
The resulting field equations are
\bea
\label{massu}
\partial_{r} m &=&  4\pi r^2 E 
\,\,\,\,\,,\\
\label{deltafi}
\partial_{r}\delta  &=& \frac{4\pi  r }{N}  
\left[ E + S^{r} \right] \,\,\,\,, \\
\label{scalareq}
\partial_{rr}^2 \phi &=& 
-\left[ \frac{2}{r} +  
\frac{2}{N} \left\{ 2\pi r (S^r - E) + 
\frac{m}{r^2} \right\} 
\right] \partial_{r}\phi  + \frac{1}{N} \left[
{\partial V(\phi)\over \partial \phi}
-2\xi\phi R \right] \,\,\,\,\,,
\eea
where $E = - T^{t}_{t}$, $S^{r} = T^{r}_{r}$ and $R$ 
are the local energy density, the radial pressure and the 
Ricci scalar, respectively,  
given by the following expressions containing no-second order derivatives: 
\bea
E &=& \frac{G_{\rm eff}}{1+16\pi\xi 
r\phi (\partial_{r}\phi) G_{\rm eff} } 
\left[ \frac{-4\xi\phi(\partial_{r}\phi)m}{r^2} 
-16\pi\xi r \phi ( \partial_{r}\phi ) G_{\rm eff}
\left( \frac{N (\partial_{r}\phi)^2}{2}
-V(\phi) -\frac{8\xi\phi(\partial_{r}\phi) N}{r} 
\right) \right] +  \nonumber \\
& & \quad \frac{G_{\rm eff}}{1+ 192\pi\xi^2 \phi^2 G_{\rm eff}}
\left[ \frac{N (\partial_{r}\phi)^2}{2} \left(1+8\xi+ 
64\pi\xi^2\phi^2 G_{\rm eff}\right) + 
4\xi\phi {\partial V(\phi)\over \partial \phi}  
+V(\phi)\left(1 - 64\pi\xi^2\phi^2 G_{\rm eff}\right) 
\right] \,\,\,\,, \\
\label{Eu}
S^{r} &=& \frac{G_{\rm eff}}{1+16\pi\xi 
r\phi (\partial_{r}\phi) G_{\rm eff} }
\left[\frac{-4\xi\phi 
(\partial_{r}\phi) m}{r^2} 
+ \frac{N (\partial_{r}\phi)^2}{2} - V(\phi) 
- \frac{8\xi\phi(\partial_{r}\phi)N}{r} \right]
\,\,\,, \\
\label{Su}   
\label{Rcur}
R &=&  
\frac{8\pi G_{\rm eff}}{1+ 192\pi\xi^2\phi^2 G_{\rm eff}}
\left[ N (\partial_{r}\phi)^2
(1 + 12\xi) + 4
V(\phi) + 12\xi\phi{\partial V(\phi)\over \partial \phi} 
\right]  \,\,\,\,.
\eea


{\it Boundary conditions and numerical methodology}.- For 
black hole configurations we demand regularity on the 
event horizon $r_h$ and for scalarons regularity at the origin 
$r=0$. This implies the following conditions for the 
fields (hereafter, the subscripts `$h$' 
stand for a quantity to be evaluated at $r_h$):
\be\label{reg}
m_h = \frac{r_h}{2} \,\,\,,\,\,\,
\delta(r_h) = \delta_h \,\,\,,\,\,\,
\phi(r_h) = \phi_h    \,\,\,\,,\,\,\,
(\partial_r \phi)_h = 
\frac{r_h \left[ (\partial_{\phi} V)_h 
- 2\xi\phi_h R_h \right]}
{4\pi r_h^2 (S_h^r - E_h) + 1}  
\,\,\,\,,
\end{equation}
where 
\bea
R_h &=& \frac{32\pi G_{\rm eff}^h}
{1 + 192\pi \xi^2 \phi_h^2 G_{\rm eff}^h}
\left[ V_h + 3\xi\phi_h (\partial_{\phi} V)_h \right] \,\,\,, \\
S_h^r - E_h &=& - \frac{G_{\rm eff}^h}
{1 + 192\pi \xi^2 \phi_h^2 G_{\rm eff}^h} 
\left[ 4\xi\phi_h (\partial_{\phi} V)_h+ 
V_h \left(1 - 64\pi\xi^2 \phi_h^2 G_{\rm eff}^h \right)
\right] - 
G_{\rm eff}^h V_h  \,\,.
\eea
 The value $\delta_h$ is fixed   
so that the desired asymptotic behavior (see below) is obtained. 
For the scalaron case, regularity at the origin $r=0$ results by 
taking the limit $r_h\rightarrow 0$ in the above regularity conditions. 
In addition to the regularity conditions, we impose  asymptotically flat
conditions on the space-time (for black holes and scalarons). These 
imply the following conditions on fields when $r \rightarrow\infty$, 
\be
m(\infty) = M_{\rm ADM}\,\,\,,\,\,\,
\delta(\infty) = 0 \,\,\,,\,\,\,
\phi(\infty) = \phi_\infty \,\,\,.
\end{equation}
Here $M_{\rm ADM}$, is the ADM-mass associated with a given configuration. Actually, the 
value $\phi_\infty$ will correspond to the local minimum (which is also a root) of 
$V(\phi)$ (see below).

For a given scalar field potential and for a fixed $\xi$ (i.e., for a given theory),
the family of SHBH configurations will be parametrized by the arbitrary free parameter 
$r_{h}$ which specifies the location of the black hole horizon. 
As explained below, $\phi_{h}$ is a shooting 
parameter rather than an arbitrary boundary value, and its value 
is determined so that the above AF 
conditions are fulfilled. Therefore for SHBH, $M_{\rm ADM}= M_{\rm ADM} (r_{h})$. 
As mentioned above, the scalarons are contained as a limiting case when $r_h\rightarrow 0$ 
(in this limit, $\phi_0$, the value of $\phi$ ar $r=0$, is the shooting parameter), 
and the corresponding configuration is characterized by a unique $M_{\rm ADM}$.

Using the above system of field equations together with the regularity 
and asymptotic conditions, we have performed a numerical analysis for 
one class of scalar field potential and for different values of 
$\xi$.
\bigskip

{\it Numerical results.} We choose the following asymmetric scalar-field 
potential leading to the desired asymptotically flat solutions:
\bea 
\label{pottwo} V(\phi) &=& \frac{\lambda}{4}\left[
(\phi-a)^2 - \frac{4(\eta_1 + \eta_2)}{3} (\phi-a) +  2\eta_1\eta_2 
\right] (\phi-a)^2 \,\,\,,
\eea
where $\lambda$, $\eta_i$ and $a$ are constants. For this class of potentials one can 
see that, for $\eta_1>2\eta_2>0$, $\phi=a$ corresponds to a local minimum,
$\phi=a+\eta_1$ is the global minimum and $\phi=a+\eta_2$ is a local maximum. 
The key point in the shape of the potential, $V(\phi)$, 
for the asymptotically flat 
solutions to exist, is that the local minimum 
$V_{\rm min}^{\rm loc}=V(a)$ is also a zero of $V(\phi)$ \cite{comment2}. 
The dynamics of scalar 
fields with such a potential has been analyzed in the past 
within the aim of studying quantum tunneling from the false vacuum 
through the true vacuum (see Ref. \cite{coleman}). A particle-mechanics analogy \cite{coleman}, 
helps to understand the existence of 
classical solutions that interpolate between the local minimum $\phi=a$
and a value, $\phi=\phi_{h,0}$, near the global minimum, and 
rolling across the local maximum. This requires a suitable shooting method in order to find 
the correct value $\phi_{h,0}$ for the scalar field to reach the local minimum 
asymptotically \cite{comment3}. We have enforced such a method here. The asymptotic behavior 
of the scalar field shows that $\phi(r) \sim a + 
{\rm const.} e^{\pm \sqrt{\lambda\eta_1\eta_2} r}/r$. 
The shooting method allows to eliminate the runaway solutions. 

Since $V(\phi)$ is not positive definite 
(we assume $\lambda>0$), the WEC is violated. On the other hand, 
taking potentials with extrema different from zero 
($V_{\rm min}^{\rm loc} \neq 0$), leads to solutions that are   
asymptotically AdS (if $V_{\rm min}^{\rm loc}<0$), 
or asymptotically de Sitter (if $V_{\rm min}^{\rm loc}>0$ and provided 
that suitable boundary conditions are imposed at the cosmological event 
horizon) with $V_{\rm min}^{\rm loc}$ acting as an effective cosmological 
constant $\Lambda_{\rm eff}$ 
(cf. Ref. \cite{japs} for an analysis with $\Lambda_{\rm eff} \neq 0$).
\begin{figure}[h]
\centerline{
\epsfig{figure=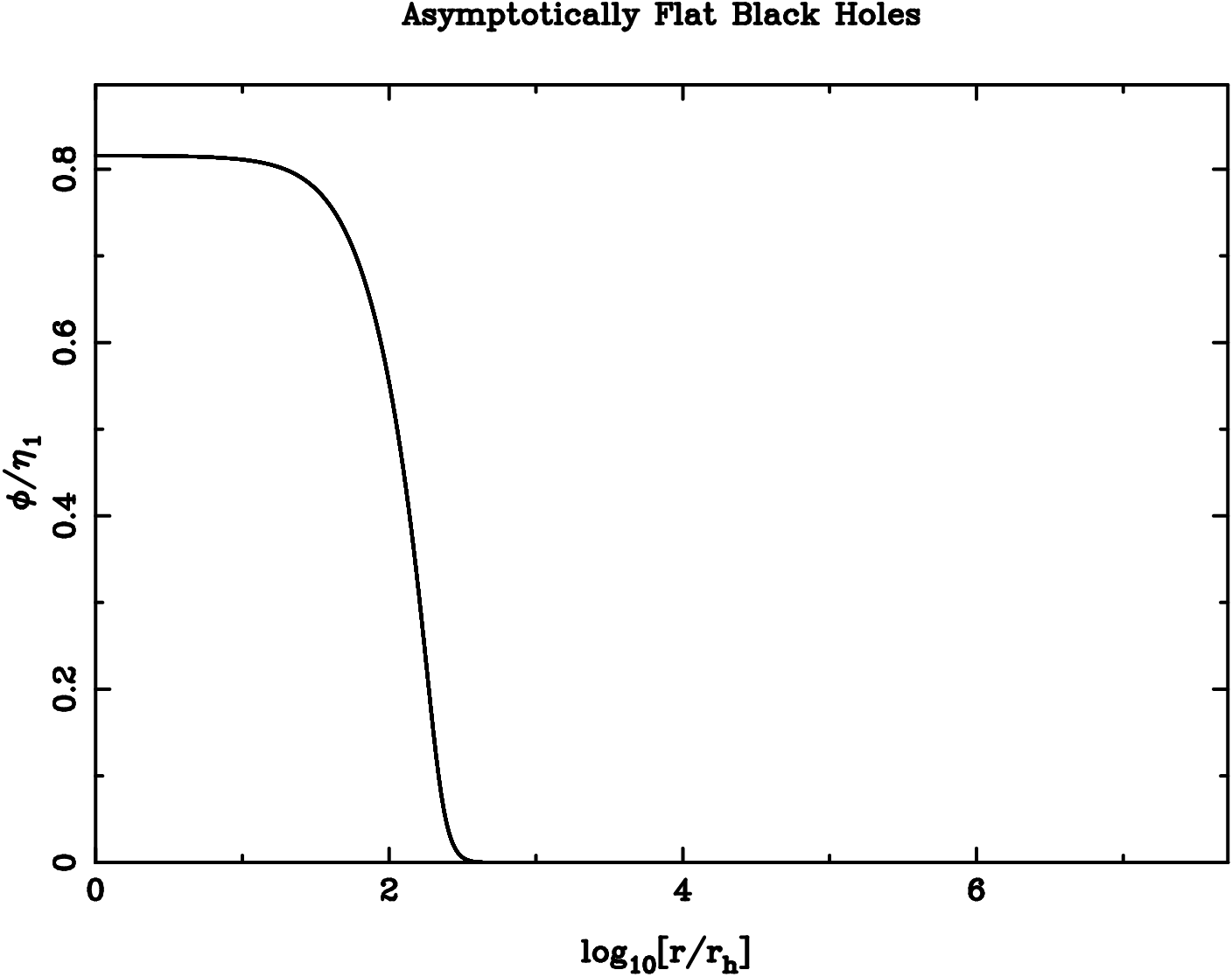,width=5.2cm,angle=-90} 
\epsfig{figure=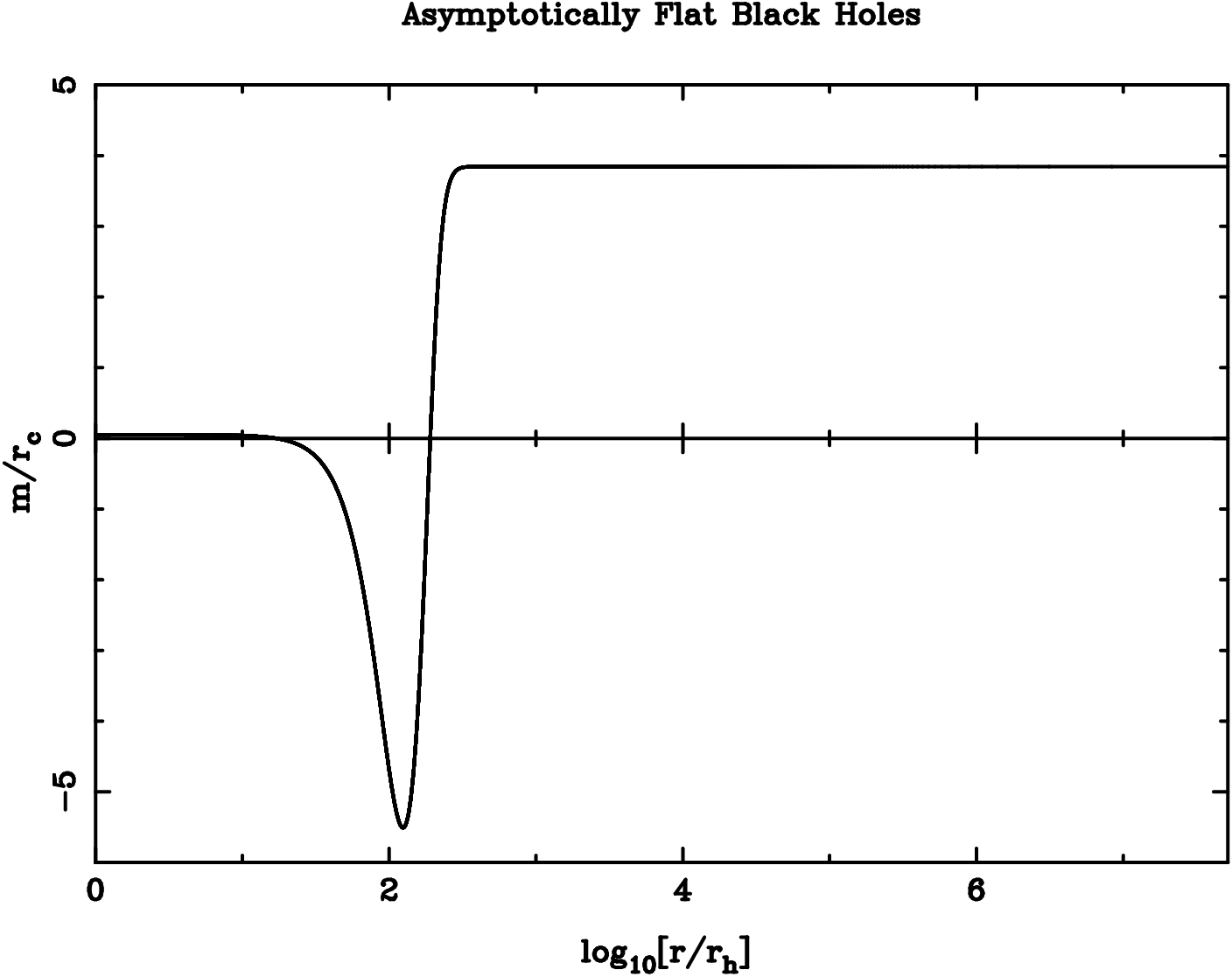,width=5.2cm,angle=-90} 
\epsfig{figure=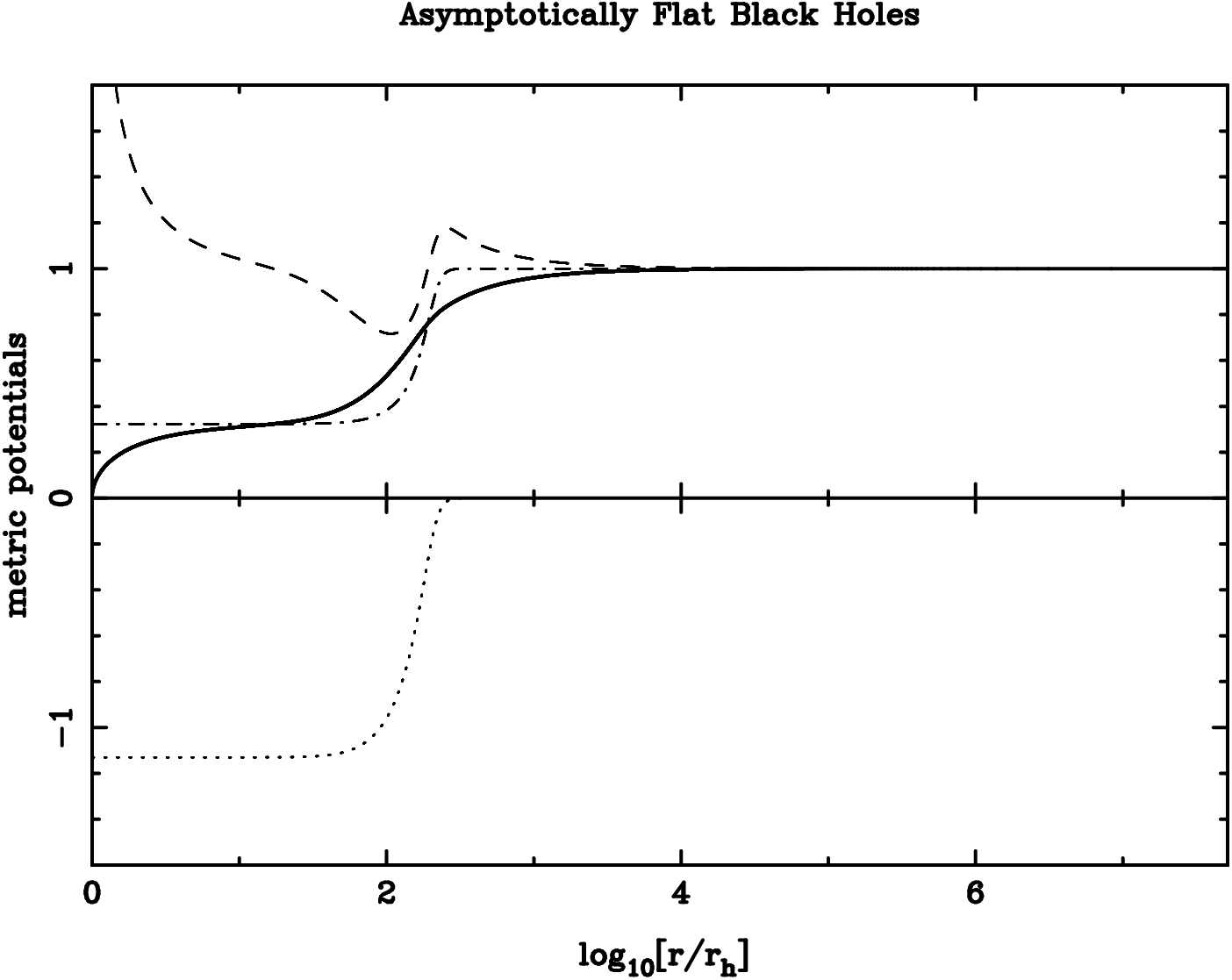,width=5.2cm,angle=-90}
}
\vspace*{0.5cm}
\caption{}
Black hole configuration constructed with $\xi=0$ 
and $V(\phi)$ as given 
by Eq. (\ref{pottwo}) with parameters $\eta_1= 0.5$, $\eta_2= 0.1$, 
$a= 0$, and  $r_h=0.1/\sqrt{\lambda}$, $\phi_h\sim 0.40786$. 
The left and middle panels depict the scalar field and the 
mass function respectively. The latter converges to 
$M_{\rm ADM}\sim 3.843 /\sqrt{\lambda}$. 
The right panel depicts the metric potentials $\sqrt{-g_{tt}}$ (solid line), 
$\sqrt{g_{rr}}$ (dashed line), $e^{\delta}$ (dash-dotted line) and 
$\delta$ (dotted line).
\label{fig1}
\end{figure}

\begin{figure}[h]
\centerline{
\epsfig{figure=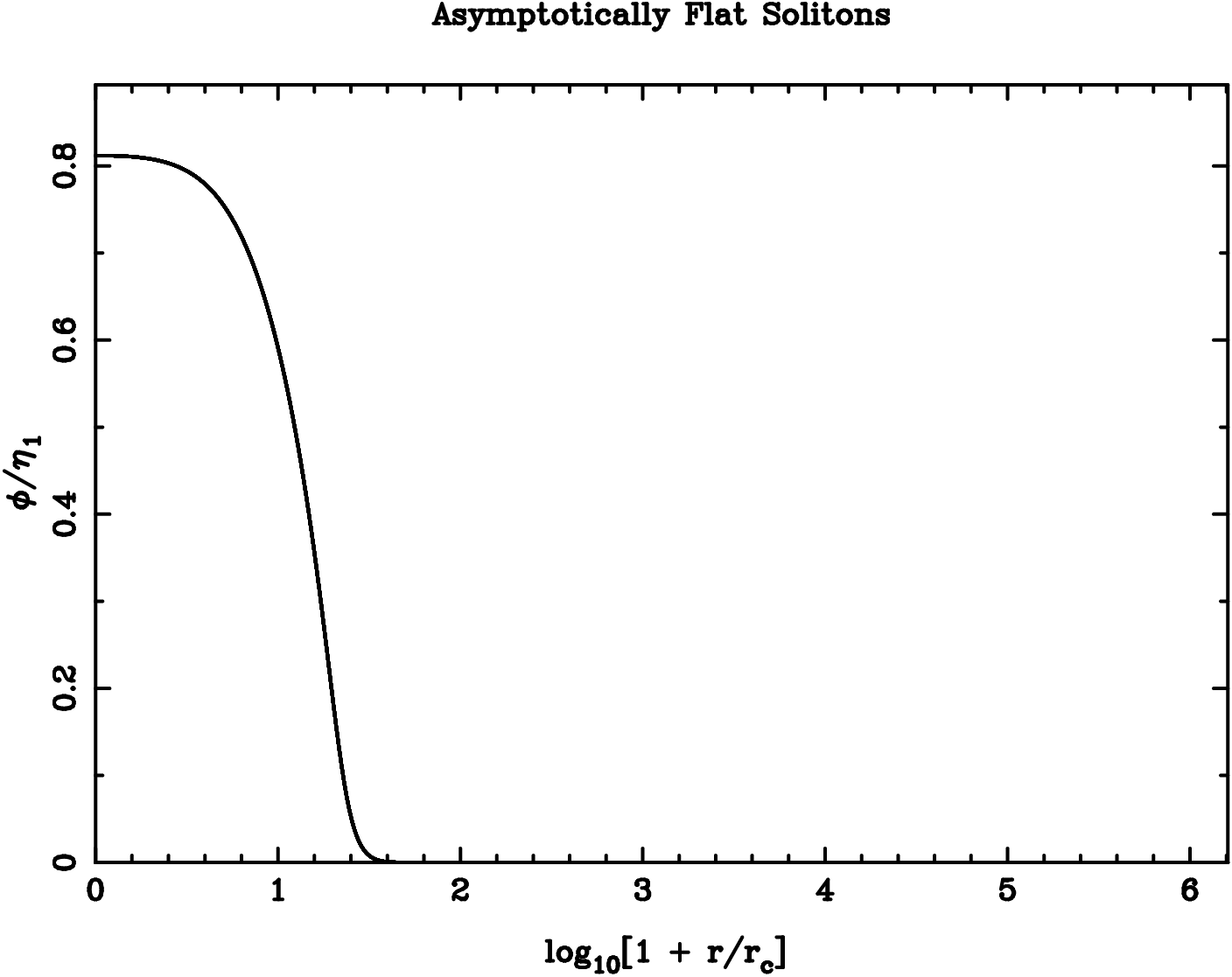,width=5.2cm,angle=-90} 
\epsfig{figure=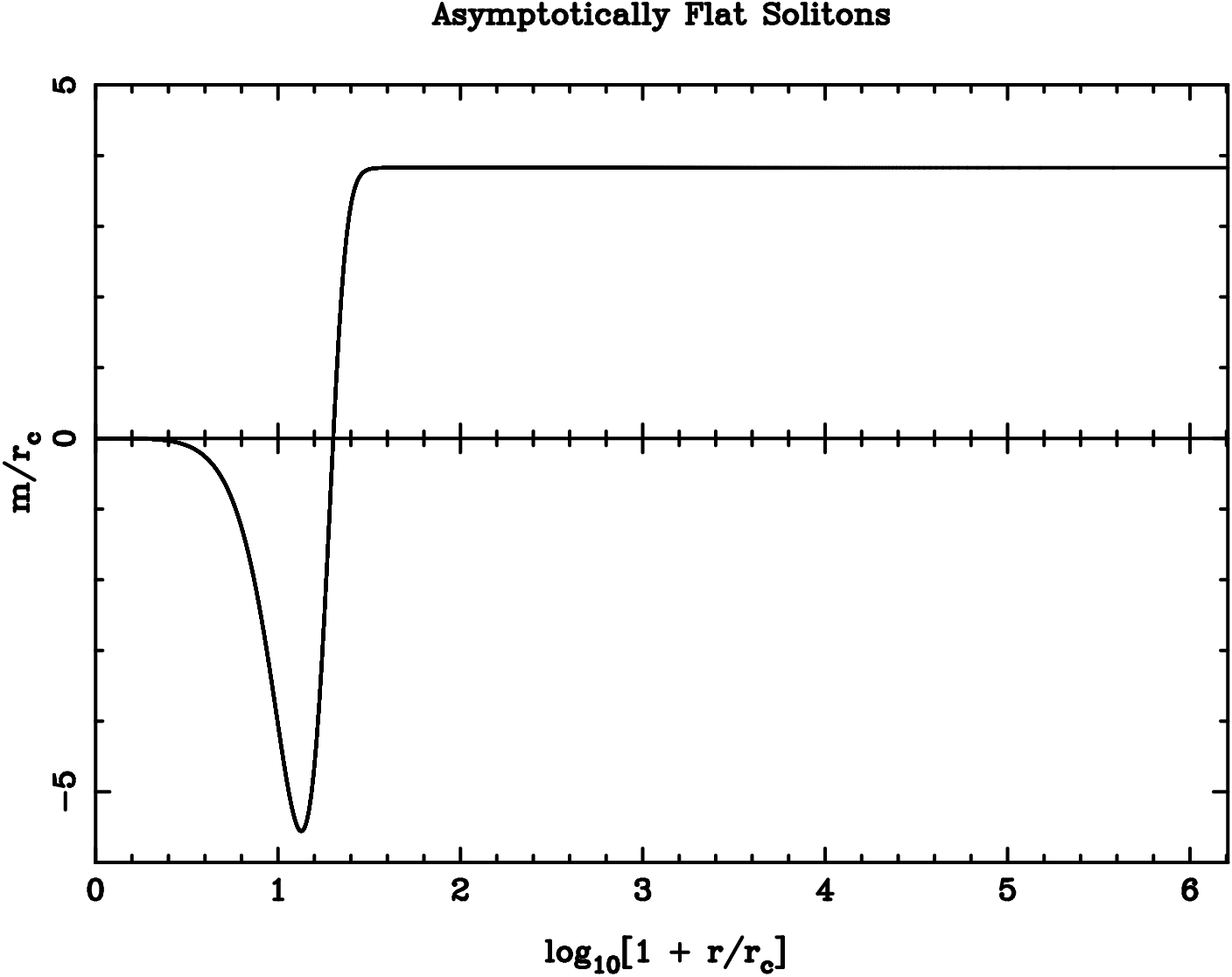,width=5.2cm,angle=-90} 
\epsfig{figure=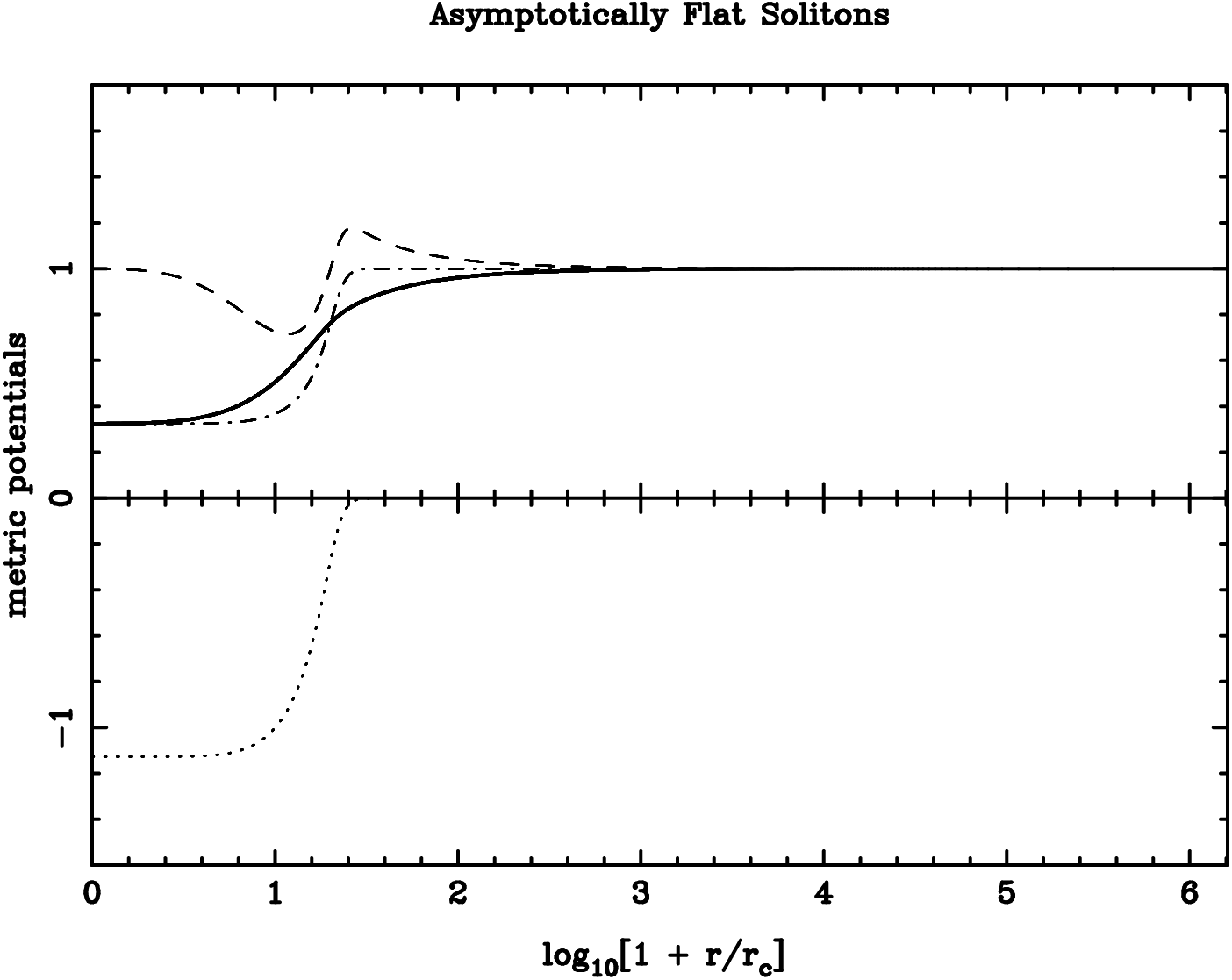,width=5.2cm,angle=-90}
}
\vspace*{0.5cm}
\caption{}
Same as Fig.\ref{fig1} for the soliton case. Here $\phi_0\sim 0.40594$ and 
$r_c=1/\sqrt{\lambda}$. Note that the solutions are globally regular, 
notably at the origin, and $M_{\rm ADM}\sim 3.827  /\sqrt{\lambda}$.
\label{fig2}
\end{figure}

For the numerical analysis, $\lambda$ determines the scale of different quantities. 
In particular $r_c=1/\sqrt{\lambda}$ has been used as a lenght scale.

{\bf Case $\xi=0$}.
Figures \ref{fig1} and \ref{fig2}, show examples of numerical 
asymptotically flat SHBH and scalaron solutions respectively, 
for the potential $V(\phi)$.  
As depicted by those figures, a generic behavior of the solution for 
$\phi$ is that it decays exponentially 
($\phi \sim {\rm const.}e^{-\sqrt{\lambda\eta_1\eta_2} r}/r$) to the 
local minimum $V_{\rm min}^{\rm loc}$ before reaching the asymptotic value at 
$\phi=a$ ($a=0$ in these examples). $\phi_{h,0}$ is the shooting 
value for which $\phi$ rolls up to $\phi=a$ asymptotically. 
The right panel of fig.\ref{fig1} (fig.\ref{fig2}) 
shows that the SHBH (scalaron) solution asymptotically 
approximates the Schwarzschild BH (Schwarzschild ``exterior'') solution with mass 
$M_{\rm ADM}^{\phi\neq 0}$ respectively.  

The ADM mass, $M_{\rm ADM} (r_{h})$, for different values of the 
parameters $(\lambda, \eta_1, \eta_2)$, turns to be larger than 
the corresponding mass of the Schwarzschild BH while keeping fixed $r_h$.
It seems therefore that the lowest bound for $M_{\rm ADM}$ is 
$M_{\rm ADM}^{\phi=0}$ which corresponds to the hairless Schwarzschild BH.
In the limit $r_h\rightarrow 0$, the lowest bound is then 
$M_{\rm ADM}=0$ which is no other but the trivial soliton of 
the Minkowski spacetime. 
The theory also admits the no-hairy solutions 
$\phi(r)= \phi_{\rm min}^{\rm global}$ 
($\phi(r)= \phi_{\rm max}^{\rm local}$), for which $\phi$ settles 
at the global minimum (local maximum) $V_{\rm min}^{\rm global}$ 
($V_{\rm max}^{\rm local}$) respectively. 
However, those solutions are not asymptotically flat, 
but rather correspond to the Schwarzschild AdS
solution (Schwarzschild dS solution) with 
$V_{\rm min}^{\rm global}$ ($V_{\rm max}^{\rm local}$) playing 
the role of a negative (positive) cosmological constant.  

Since $M_{\rm ADM}^{\phi\neq 0}(r_h) > M_{\rm ADM}^{\phi=0}(r_h)$, i.e.,
$M_{\rm ADM}(r_h)$ is bounded from below     
(while keeping fixed $r_h$ which in these coordinates is equivalent to 
fixing the area, $A_h$, of the horizon) with $M_{\rm ADM}^{\phi=0}$ corresponding to the Schwarzschild black hole, 
heuristically one expects that the hairy BH configurations found here  
with fixed AF boundary conditions are unstable, since 
the stable ones would correspond to those with the lowest 
$M_{\rm ADM}(r_h)$ (cf. Ref. \cite{sudwald}). Such a behavior is a signature 
of the unstable nature of the SHBH configurations 
which is confirmed by a perturbation analysis (similar arguments 
can be applied to the scalarons to exhibit their unstable nature). 

{\bf Case $\xi\neq 0$}. 
After performing numerous experiments for the minimal coupling case 
$\xi=0$, we also found (for some set of values $\xi> 0$) that asymptotically 
flat hairy black holes and scalarons exist for the NMC case 
 as well (the resulting figures 
are qualitatively similar to Figs.\ref{fig1} and \ref{fig2}).
 However, a more exhaustive analysis is to be performed in this case. 
For instance, the stability in the NMC is an issue that remains to 
be investigated. This will be the object of a forthcoming paper 
in which we also plan to provide a full sample of black hole 
and soliton configurations, and to analyze more carefully the quantitative 
behavior of $M_{\rm ADM}= M_{\rm ADM} (r_{h},\xi)$ 
for a larger range of $\xi$ \cite{ns}.

\bigskip

\begin{figure}[h]
\centerline{
\epsfig{figure=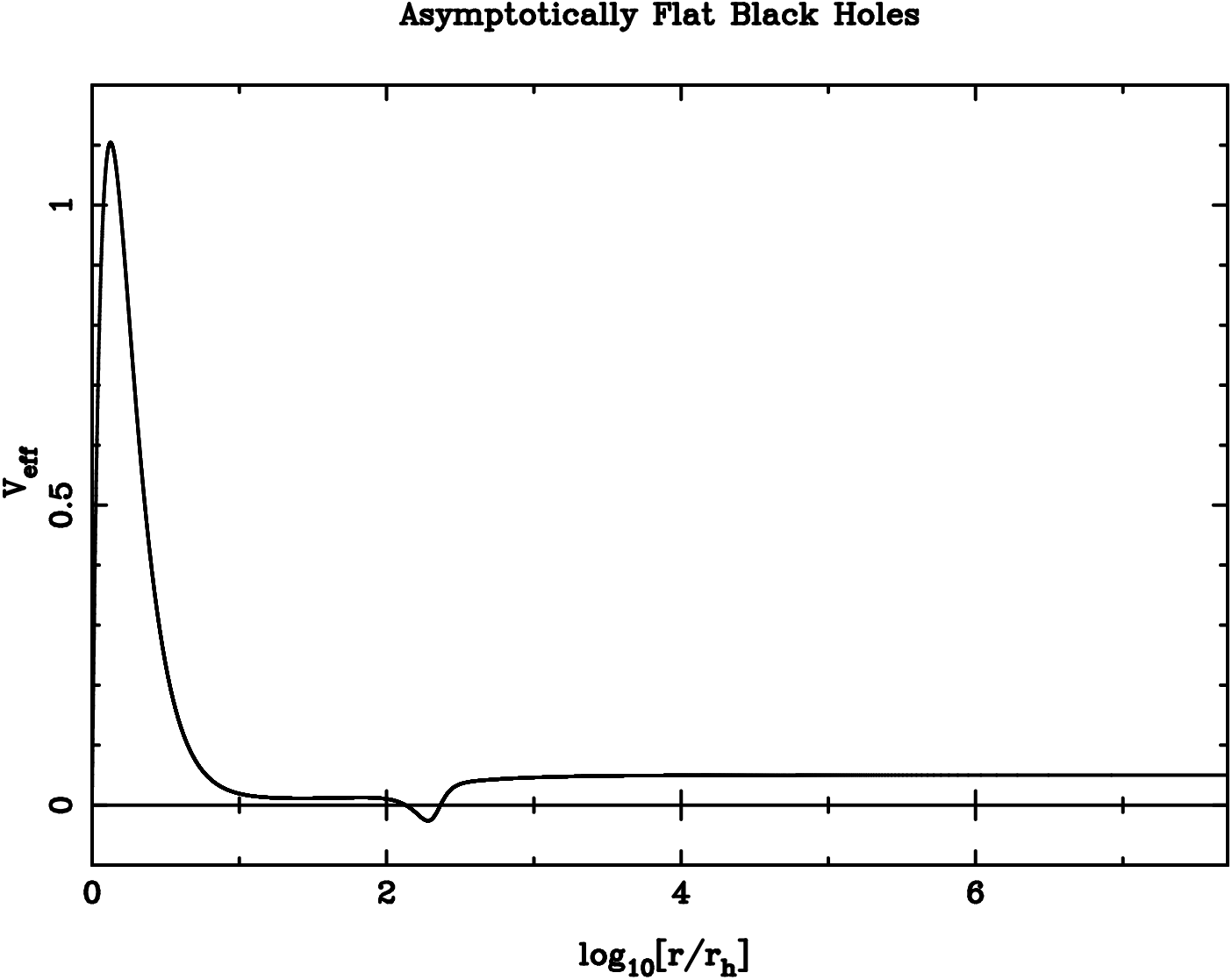,width=7cm,angle=-90} 
\epsfig{figure=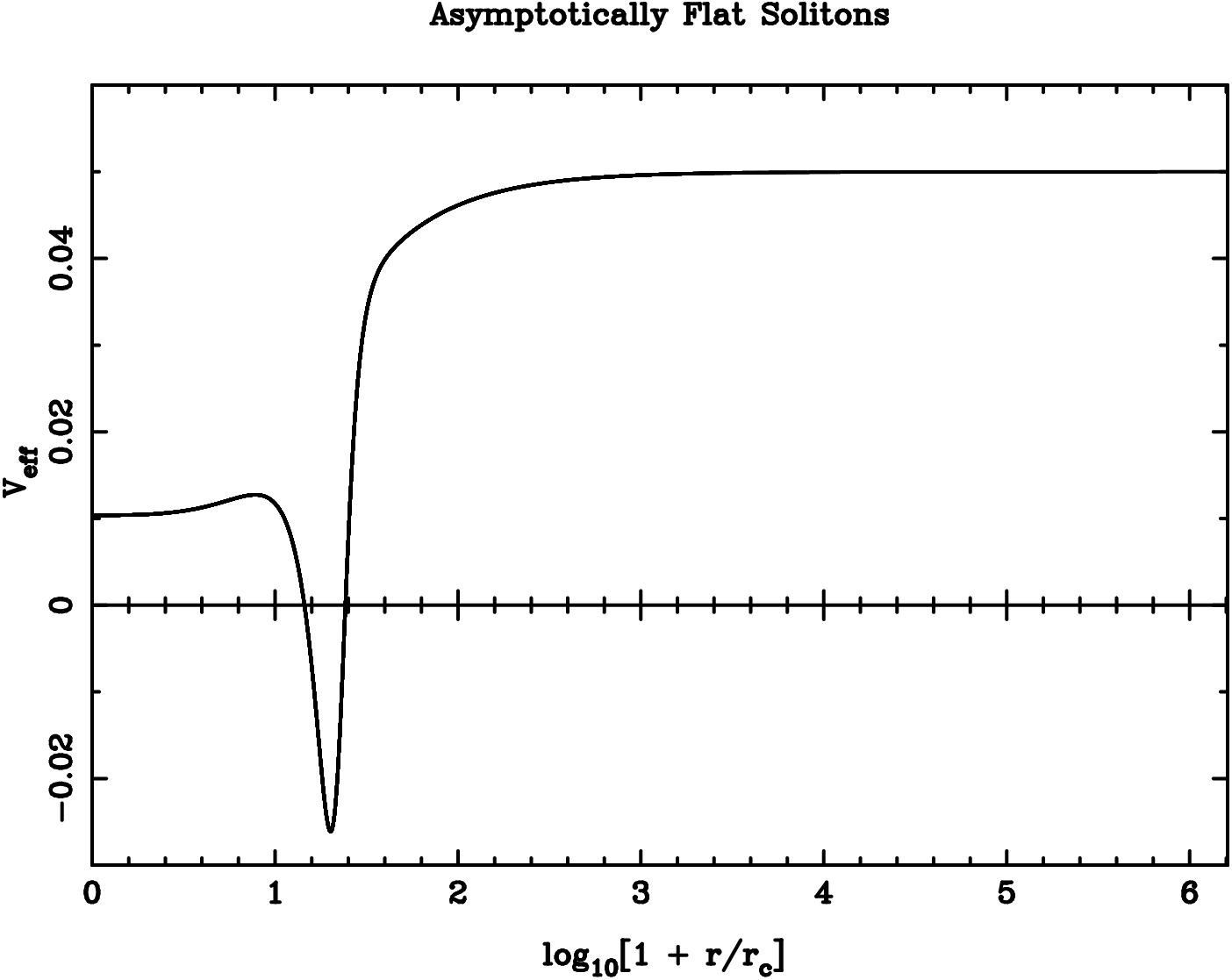,width=7cm,angle=-90} 
}
\vspace*{0.5cm}
\caption{}
The potential $V_{\rm eff}$ corresponding to the SHBH solution of fig.1 
(left panel) and to the soliton solution of fig.2 (right panel).
\label{fig3}
\end{figure}

{\it Linear stability analysis.-} We now turn to the question of
the stability of the asymptotically flat space-times presented here. 
For simplicity, the following analysis is limited to the case $\xi=0$.  
We consider only spherically symmetric time-dependent linear
perturbations to the spherically symmetric static metric
(\ref{RGMS}) and the scalar field configuration discussed in the
previous section. These perturbations will be parametrized by
\bea
\tilde\phi(t,r) &=& \phi(r) + \phi_1(t,r) \,\,\,, \\
\tilde g_{tt}(t,r) &=& g_{tt}(r) (1 - h_0(t,r)) \,\,\,, \\
\tilde g_{rr}(t,r) &=& g_{rr}(r) (1 + h_2(t,r)) \,\,\,,
\eea
where $g_{tt}(r) = - N e^{2\delta}$, $g_{rr}(r) = N^{-1}$,
$\phi(r)$ represent the non-perturbed
solution and $\phi_1(t,r)$, $h_0(t,r)$, $h_2(t,r)$ denote a small
perturbation to the non-perturbed solution. It is straightforward to
obtain the metric perturbations as a function of the scalar field
perturbation
\be
h_2 = 8\pi r (\partial_r \phi) \phi_1 \,\,\,,\,\,\,
\partial_r h_0 = \partial_r h_2 - 16\pi r (\partial_r \phi)
(\partial_r \phi_1) \,\,\,,
\end{equation}
and the linear Schr\"odinger type equation for the scalar field perturbation
\be\label{perteq}
-\partial_{r_*r_*}^2 \psi + V_{\rm eff}^*[r_*] \psi = 
-\partial_{tt}^2 \psi  \,\,\,,\,\,\,
\frac{dr_*}{dr} = \frac{e^{-\delta}}{N} \,\,\,,
\end{equation}
where we have used $\psi \equiv r \phi_1$, $r_*(r)$ is the ``tortoise'' 
coordinate, and the effective potential is given by
\bea
\label{poteff}
V_{\rm eff}(r) &=& N e^{2\delta} \left[ \frac{N}{r}
\left\{ \partial_r \delta + \frac{\partial_r N}{N} \right\}
- 8\pi r N (\partial_r \phi)^2 \left\{ \partial_r \delta +
\frac{\partial_r N}{N} + \frac{1}{r} \right\} +
16\pi r (\partial_r \phi)
{\partial V \over \partial \phi} +
{\partial^2 V \over \partial \phi^2} \right] \,\,\,.
\eea
Figure \ref{fig3} depicts $V_{\rm eff}$ for the SHBH (left panel) and 
soliton (right panel) respectively.

\begin{figure}[h]
\centerline{
\epsfig{figure=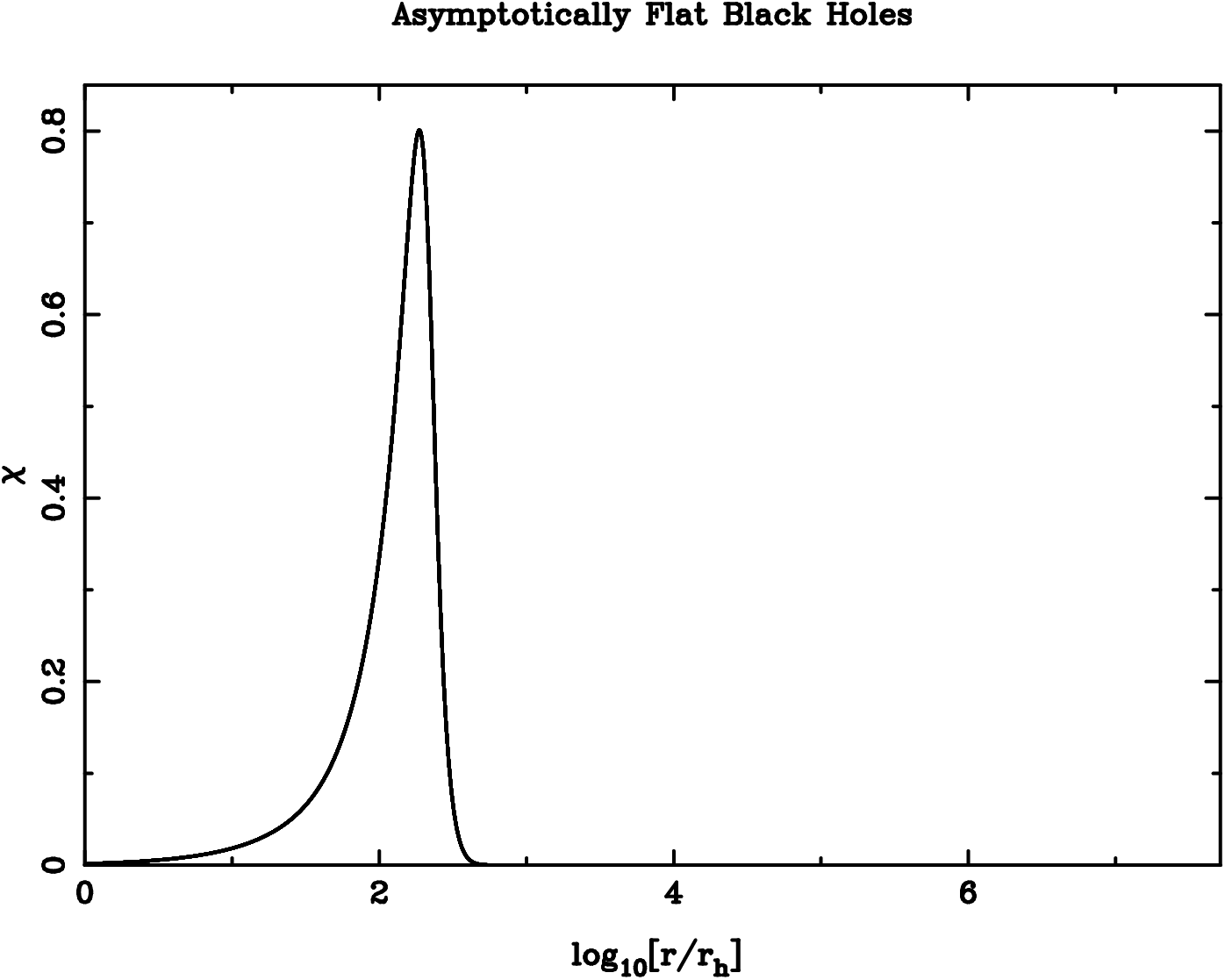,width=7cm,angle=-90} 
\epsfig{figure=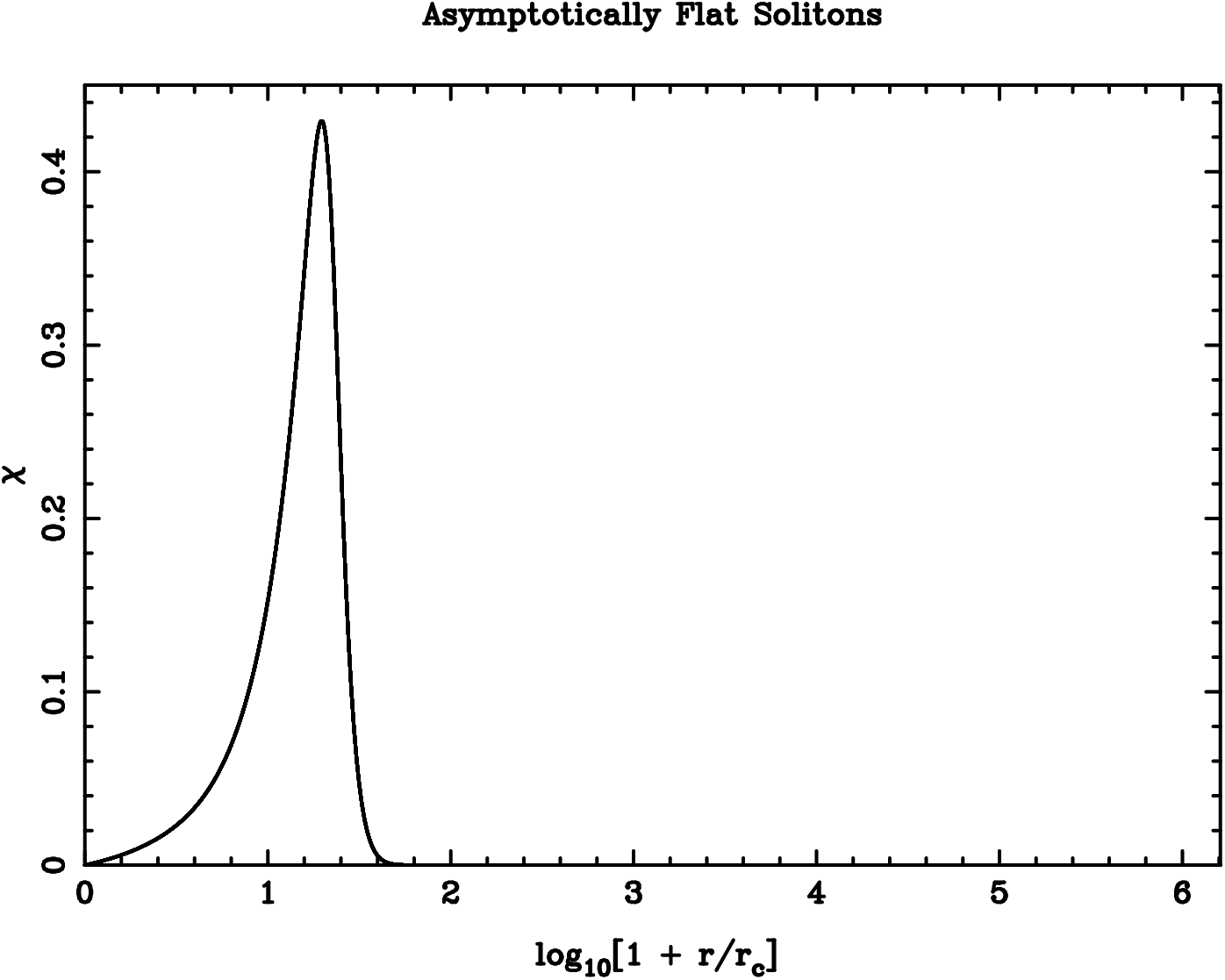,width=7cm,angle=-90} 
}
\vspace*{0.5cm}
\caption{}
The mode $\chi$ corresponding to the perturbation of the SHBH solution of fig.1 
(left panel) with $\sigma^2\sim -0.00241$ 
and the soliton solution of fig.2 (right panel) 
with $\sigma^2\sim -0.00243$ respectively.
\label{fig4}
\end{figure}

One can seek mode perturbations $\psi(r)= \chi(r)e^{\imath \sigma t}$, 
so that the Equation (\ref{perteq}) writes 
$\left(-D^aD_a + V_{\rm eff}\right) \chi = \sigma^2 \chi $ where 
$D_a$ is the derivative operator associated with an auxiliary metric 
of a manifold $M$. According to the theorem proved in \cite{wald}, 
a sufficient condition for the static and spherically 
symmetric configurations to be unstable, is that the 
operator $A=-D^aD_a + V_{\rm eff}$ 
be negative in the Hilbert space $L^2(M)$. This theorem can be 
easily implemented when the background (static) solution is given 
analytically. For instance, one can see explicitly, 
that the Schwarzschild solution 
$\delta(r)=0$, $N= 1 - 2M_{\rm ADM}^{\phi=0}/r$ 
is stable within these kind of perturbations and within this 
class of potentials, since 
$V_{\rm eff}(r)= N\left[\frac{2M_{\rm ADM}^{\phi=0}}{r^3} 
+ \lambda \eta_1 \eta_2 \right] > 0$ 
in the region $r_h\leq r \leq \infty$. 
A result, which is heuristically confirmed by the fact that 
$M_{\rm ADM}^{\phi=0}$ is a lower bound 
within the set of configurations with fixed $r_h$ \cite{ns}. 
However, when the background solution is given numerically the implementation 
of the theorem is less trivial since it requires the use of a suitable 
square integrable vector $\Psi$ in $L^2(M)$ to compute $<\Psi,A\Psi>$. 
Instead, we have opted to explicitly solve the associated Schr\"odinger-like 
equation of Eq. (\ref{perteq}) with respect to the 
original $r$ coordinate to find a ``bound state'' with negative $\sigma^2$.  
Imposing regularity at $r_h$ (origin), i.e., $\chi|_{r_h,0}=0$, 
and demanding that $\chi(r)\rightarrow 0$ 
for $r\rightarrow \infty$ we found $\sigma^2\sim -0.00241$ and 
$\sigma^2\sim -0.00243$ for the SHBH and scalarons solutions 
respectively (see Fig. \ref{fig4}). This shows that 
the modes $\psi(r)= \chi(r)e^{\imath \sigma t}$ can grow unboundedly with time, 
and therefore leading to the conclusion that 
the SHBH and soliton configurations are unstable.

{\it Conclusions.} We have found black hole solutions 
that support a nontrivial scalar hair in spherical, static and 
asymptotically flat spacetimes. Such solutions are perhaps the 
first regular ones throughout the static region 
(from the horizon to spatial infinity) with an explicit scalar field
potential \cite{hairy}. In the limit where 
$r_h\rightarrow 0$, regular scalar solitons (scalarons) are obtained. In the 
minimal coupling case $\xi=0$, the solutions turn to be unstable 
with respect to linear-spherical perturbations and their ADM mass 
turns to be larger than the corresponding Schwarzschild black hole
while keeping fixed $r_h$. 
Therefore, these are to be regarded as 
weak counterexamples to the no-scalar-hair conjecture. For  
nonminimal couplings $\xi \neq 0$, similar solutions are found, 
however their stability remains to be investigated. 
Another analysis to be performed are the collapse of these hairy black holes 
violating WEC 
and the possibility of having hairy black holes with nodes 
in the scalar field using higher order potentials. 
An interesting outcome of the above results is that it 
opens the possibility for a further study of the solutions 
in the context of the isolated horizon formalism (cf. Ref. \cite{ash}).
\bigskip

We are indebted to A. Corichi, D. Sudarsky and T. Zannias for discussions, 
and to E. Ay\'on for providing some references. We particularly thank 
O. Sarbach for his comments and suggestions on the stability analysis. 
U.N. would like to acknowledge partial support from SNI. 
M.S. acknowledges partial support from Conacyt grant No. 32551-E and 
DGAPA-UNAM grants No. IN112401 and No. IN122002. 



\begin{thebibliography}{}

\bibitem{ruffini}
R. Ruffini and J. A. Wheeler, Phys. Today {\bf 24}(1), 30 (1971).

\bibitem{israel}
W. Israel, \Journal{\PR}{164}{1776}{1967};
\Journal{\CMP}{8}{245}{1971}; B. Carter,
\Journal{\PRL}{26}{331}{1971}; R. Wald,
\Journal{\ibid}{26}{1653}{1971}; D. C. Robinson,
\Journal{\ibid}{34}{905}{1977}; P. Q. Mazur,
\Journal{\JPA}{15}{3173}{1982}; \Journal{\PLA}{100}{341}{1984}.

\bibitem{chase}
J. E. Chase, \Journal{\CMP}{19}{276}{1970}.

\bibitem{bekenstein}
J. D. Bekenstein, \Journal{\PRD}{5}{1239}{1972}.

\bibitem{hartle}
J. B. Hartle, in {\it Magic Without Magic}, edited by J. Klauder
(Freeman, San Francisco, 1972).

\bibitem{teitelboim}
C. Teitelboim, \Journal{\PRD}{5}{2941}{1972}.

\bibitem{eym}
P. Bizon, \Journal{\PRL}{64}{2844}{1990}; M. S. Volkov, and D. V. Gal'tsov, 
Sov. J. Nucl. Phys., {\bf 51}, 1171 (1990); H. P. Kunzle, and A. K. M. Masood-ul-Alam, 
\Journal{\JMP}{31}{928}{1990} 

\bibitem{eymh}
B. R. Greene, S. D. Mathur, C. M. O'Neill, \Journal{\PRD}{47}{2242}{1993}.

\bibitem{eymd}
G. Lavrelashvili, and D. Maison, Nucl. Phys. {\bf B 410}, 407 (1993)

\bibitem{es}
P. Bizon, and T. Chmaj, \Journal{\PLB}{297}{55}{1992}; M. Heusler, 
S. Droz, and N. Straumann, \Journal{\PLB}{268}{371}{1991}; {\it idem} 
\Journal{\PLB}{271}{61}{1991}; \Journal{\PLB}{285}{21}{1992}.
  
\bibitem{review}
M. S. Volkov, and D. V. Gal'tsov, 
Phys. Rept., {\bf 119}, 1 (1999).

\bibitem{comment1}
In theories such as EYMH or EYMD where scalar field hair exists,
this one disappears when the non-Abelian gauge fields vanishes.

\bibitem{sud1}
D. Sudarsky, \Journal{\CQG}{12}{579}{1995}. 

\bibitem{bekenstein2}
J. D. Bekenstein, \Journal{\PRD}{51}{R6608}{1995}.

\bibitem{wald2}
R. M. Wald, General Relativity and Gravitation, 
University of Chicago Press, Chicago (1984). 

\bibitem{zannias}
T. Zannias, \Journal{\JMP}{36}{6970}{1995}.

\bibitem{nomin} 
B. C. Xanthopoulos, and T. Zannias, \Journal{\JMP}{32}{1875}{1991}; 
A. Saa, \Journal{\PRD}{53}{7377}{1996}; 
\Journal{\JMP}{37}{2349}{1996}; A. E. Mayo, and J. D. Bekenstein, 
\Journal{\PRD}{54}{5059}{1996}; E. Ay\'on-Beato \Journal{\CQG}{19}{5465}{2002} 

\bibitem{BMBB}
J. D. Bekenstein, Ann. Phys. (NY), {\bf 82} 535 (1974); N. Bocharova, K. Bronikov, and 
V. Melnikov, Vestn. Mosk. Univ. Fiz. Astron., {\bf 6} 706 (1970)

\bibitem{sudzann}
D. Sudarsky, and T. Zannias, \Journal{\PRD}{58}{087502}{1998}.

\bibitem{japs}
T. Torii, K. Maeda, and M. Narita, \Journal{\PRD}{64}{044007}{2001}; 
E. Winstanley, (pre-print gr-qc/0205092)

\bibitem{sudja}
D. Sudarsky, and J. A. Gonz\'alez, \Journal{\PRD}{67}{024038}{2003}

\bibitem{comment2}
Taking potentials with a local maximum coinciding with a zero leads to 
scalar field solutions that oscillate around that maximum and which correspond
to ``quasi-asymptotically'' flat spacetimes. That is, asymptotically 
$\phi(r) \sim a + A \sin(\sqrt{\lambda\eta_1\eta_2} r + \alpha)/r$ 
($A,\alpha$ are constants), and the spacetimes asymptotically 
resemble to the Minkowski spacetime but their ADM-mass is 
not well defined [i.e., asymptotically $m(r) \sim {\rm const.} + 
A^2 \pi \sqrt{\lambda\eta_1\eta_2} 
\sin[2(\sqrt{\lambda\eta_1\eta_2} r + \alpha)]$.

\bibitem{coleman}
S. Coleman, \Journal{\PRD}{15}{2929}{1977}; 
\Journal{\PRD}{16}{1248}{1977}.

\bibitem{comment3}
One could expect that the shooting values $\phi_{h,0}$ are not unique. 
However, this would happen only if the ``initial'' data  
$\partial_r\phi|_{h,0}$ were not constrained by the regularity 
conditions at the horizon (origin). For instance, using the 
particle analogy of \cite{coleman}, one could think that there are also shooting values 
on the right (left) of the global (local) 
maximum of $-V(\phi)$ such that with a suitable initial 
``velocity'' $\partial_r\phi|_{h,0} \neq 0$ the scalar field will roll-up-down-up 
(roll up) asymptotically 
to the local maximum of $-V(\phi)$. However, in the soliton case, the initial 
``velocity'' $\partial_r\phi|_0 = 0$ (by regularity), and therefore, it seems that 
there is only one $\phi_0$ (with $\partial_r\phi|_0=0$) that 
provides the desired shooting value. For the black hole, 
$\partial_r\phi|_h= f(r_h,\phi_h)$ is not zero [see (\ref{reg})]. 
It could be then possible, that 
by solving the four order polynomial $\partial_r\phi|_h= f(r_h,\phi_h)$ 
for $\phi_h$ in terms of $r_h$, $\partial_r\phi|_h$ [and the parameters of 
$V(\phi)$], one can find another successful shooting value (besides the one 
found here) now using $\partial_r\phi|_h$ as shooting parameter
(with $\partial_r\phi|_h$
large enough for $\phi$ to roll through the corresponding hills of $-V(\phi)$ ) 
such that another real-valued root $\phi_h$ of the above polynomial exists. 
This certainly will depend on the values of the parameters $\eta_1$ and $\eta_2$ and 
$r_h$.

\bibitem{sudwald}
D. Sudarsky, and R. M. Wald, \Journal{\PRD}{46}{1453}{1992}.

\bibitem{wald}
R. M. Wald, \Journal{\JMP}{33}{248}{1992}.

\bibitem{ns}
U. Nucamendi and M. Salgado, (in preparation)

\bibitem{hairy}
At the time of preparation of this paper we were pointed out the 
following references:
O. Bechmann and O. Lechtenfeld, \Journal{\CQG}{12}{1473}{1995}; 
H. Dennhardt and O. Lechtenfeld, \Journal{\IJMPA}{13}{741}{1998}
, where  
``analytic'' scalar-hairy black hole solutions are found with implicit 
scalar-field potentials. 
In a similar fashion K.A. Bronnikov, and G.N. Shikin, Grav. Cosmol. 
{\bf 8}, 107 (2002), have constructed some sort of ad-hoc 
scalar solitons and hairy black hole solutions.

\bibitem{ash}
A. Ashtekar, A. Corichi, and D. Sudarsky, \Journal{\CQG}{18}{919}{2001}.


\end{thebibliography}
\end{document}